\documentstyle[11pt,amssymb]{article}

\makeatletter
\@addtoreset{equation}{section}
\makeatother

\def\dalemb#1#2{{\vbox{\hrule height .#2pt
        \hbox{\vrule width.#2pt height#1pt \kern#1pt
                \vrule width.#2pt}
        \hrule height.#2pt}}}

\def\cL{{\cal L}}

\def\0{{\sst{(0)}}}
\def\1{{\sst{(1)}}}
\def\2{{\sst{(2)}}}
\def\3{{\sst{(3)}}}
\def\4{{\sst{(4)}}}
\def\5{{\sst{(5)}}}
\def\6{{\sst{(6)}}}
\def\7{{\sst{(7)}}}
\def\8{{\sst{(8)}}}
\def\n{{\sst{(n)}}}

\def\CS{{\cal S}}

\def\ep{\epsilon}
\def\td{\tilde}

\def\half{{\textstyle{1\over2}}}
\def\qu{{\textstyle{1\over 4}}}

\let\a=\alpha \let\b=\beta  \let\d=\delta 
    \let\k=\kappa
 \let\m=\mu \let\n=\nu  \let\r=\rho
\let\s=\sigma \let\t=\tau  \let\f=\phi  
  \let\D=\Delta  
    
 \let\W=\Omega     
\let\la=\label \let\ci=\cite \let\re=\ref
\let\se=\section   
\def\nn{\nonumber} \def\bd{\begin{document}} \def\ed{\end{document}}
\def\ds{\documentstyle} \let\fr=\frac \let\bl=\bigl \let\br=\bigr
\let\Br=\Bigr \let\Bl=\Bigl 
\let\bm=\bibitem
\let\na=\nabla
\let\pa=\partial \let\ov=\overline 
\newcommand{\be}{\begin{equation}} 
\newcommand{\ee}{\end{equation}}
\def\ba{\begin{array}}
\def\ea{\end{array}}
\def\ft#1#2{{\textstyle{{\scriptstyle #1}\over {\scriptstyle #2}}}}
\def\fft#1#2{{#1 \over #2}}
\def\del{\partial}
\def\sst#1{{\scriptscriptstyle #1}}
\def\oneone{\rlap 1\mkern4mu{\rm l}}
\def\ie{{\it i.e.\ }}
\def\via{{\it via}}
\def\semi{{\ltimes}}
\def\str{{\rm str}}
\def\Dm{{{D_{\sst{max}}}}}

\newcommand{\hsp}{\hspace{0.5cm}}

\newcommand{\ho}[1]{$\, ^{#1}$}
\newcommand{\hoch}[1]{$\, ^{#1}$}
\newcommand{\bea}{\begin{eqnarray}} 
\newcommand{\eea}{\end{eqnarray}} 
\newcommand{\ra}{\rightarrow}
\newcommand{\lra}{\longrightarrow}
\newcommand{\Lra}{\Leftrightarrow}
\newcommand{\ap}{\alpha^\prime}
\newcommand{\bp}{\tilde \beta^\prime}
\newcommand{\tr}{{\rm tr} }
\newcommand{\Tr}{{\rm Tr} } 
\newcommand{\NP}{Nucl. Phys. }
\newcommand{\upenn} {\it DAMTP, University of Cambridge, 
CMS, Wilberforce Road, Cambridge. CB3 0WE}
\newcommand{\spin}{\it Spinoza Institute, University of Utrecht,
Postbus 80.195, 3508 TD Utrecht}

\newcommand{\auth}{Marika Taylor-Robinson\hoch{\dagger}}

\begin{document}
\begin{flushright}
\hfill{\bf hep-th/0110142}\\
\hfill{SPIN-2001/21} \\
\hfill{October 2001}\\
\end{flushright}

%\vspace{15pt}

\begin{center}
{ \large {\bf Higher-dimensional Formulation of Counterterms}}

\vspace{10pt}
\auth

\vspace{15pt}

{\spin}

\vspace{12pt}

\underline{ABSTRACT}
\end{center}
\noindent
It is by now well established that divergences of the on-shell action 
for asymptotically AdS solutions can be cancelled by adding covariant local
boundary counterterms to the action. Here we show that although one can
still renormalise the action for asymptotically $AdS_p \times S^q$ solutions
using local boundary counterterms the counterterm action is not covariant
since the conformal boundary is degenerate. 
Any given counterterm action is defined with respect to
specific coordinate frame and gauge choices.

{\vfill\leftline{}\vfill
%\vskip 5pt
\footnoterule
{\footnotesize  \hoch{\dagger} M.TaylorRobinson@phys.uu.nl }
\vskip  -12pt} \vskip 14pt

\pagebreak
\setcounter{page}{1}

\tableofcontents
\addtocontents{toc}{\protect\setcounter{tocdepth}{2}}
\newpage

\se{Introduction}
\noindent
The correspondence between supergravity (string)
theories in AdS backgrounds and 
their dual conformal field theories has by now passed many tests and 
produced a number of interesting insights into strong coupling behaviour of
field theory \ci{m}, \ci{w}, \ci{gkp}, \ci{a}. 
It has also provided a partial solution to a long standing
problem in relativity: how to define finite masses and charges for a given 
asymptotically AdS metric without resorting to an ill-defined
background subtraction procedure. 

The method of holographic renormalisation has been developed systematically 
for asymptotically AdS solutions \ci{bk}, \ci{kls}, \ci{ejm}, \ci{hs}, 
\ci{hss}, \ci{s}, \ci{bzs}, \ci{mu}. A finite renormalised
action can be constructed as a functional of boundary data for the bulk fields
which are sources of operators in the dual field theory. One can obtain from
this action the renormalised correlation functions by functional 
differentiation.

AdS backgrounds in string theory appear as backgrounds of the form
$AdS_p \times X^q$ where $X^q$ is an Einstein manifold of positive curvature
whose symmetries relate to the $R$ symmetries of the dual theory. So far
all holographic renormalisation has been carried out at the level of the 
$p$-dimensional gauged supergravity action obtained by dimensionally reducing
on $X^q$.

\bigskip

Here we attempt to extend the program of holographic renormalisation to the
$(p+q)$-dimensional action. There are a number of motivations for this. 
It would be interesting to understand how the full higher dimensional spacetime
is reconstructed from field theory data. Also 
the Kaluza-Klein reduction is complex and explicitly matching higher to lower
dimensional fields is not easy even for quite simple solutions, such as BPS
brane distributions. There are a number of solutions which are known in either
higher or lower dimension but which have not been explicitly lifted or reduced.
One example is the GPPZ flow in five dimensions \cite{gppz} 
which should lift to a Polchinski-Strassler solution \cite{ps}. 
Part of the lift was carried out in \cite{pw} but
the full ten-dimensional set of fields is still not known. So it would be
useful, purely as a calculational tool, to be able to calculate the 
renormalised action, mass and so on in both dimensions.

One should be able to extend the techniques of holographic renormalisation to
supergravity in any background with a dual field theory description for which
UV divergences can be regulated and removed by a suitable renormalisation
scheme. In particular, solutions of gauged supergravity theories such as those
considered in \cite{mn} should admit holographic renormalisation, even though
they are not asymptotically AdS. Such backgrounds share a significant feature
with $AdS_p \times X^q$ backgrounds: the conformal boundaries are degenerate. 
In the latter case, this degeneracy is manifest since the boundary is the
product of some $(p-1)$-dimensional manifold and (the collapse of $X^q$ to)
a point. 

Degenerate boundaries were explored in \cite{t} and further understanding 
the implications of such degeneracy is the main topic of this paper. The main
result is the following. Although we can renormalise the action using a local 
boundary counterterm action, the counterterms cannot be written in terms of
covariant quantities intrinsic to the boundary. This means that for a given
counterterm action we will also have to specify particular coordinate 
and gauge choices. We should emphasise that nonetheless we can 
still renormalise the action in a well-defined and systematic procedure. 

To demonstrate these points we will first consider divergences and 
counterterms for asymptotically $AdS_3 \times S^3$ spacetimes. In \S 3 
we then discuss the physically more interesting case of asymptotically 
$AdS_5 \times S^5$ spacetimes and demonstrate explicitly how the renormalisation
procedure works from the higher-dimensional perspective. 

\se{Divergences and counterterms for $AdS_3 \times S^3$ spacetimes}

Three-dimensional counterterms have been
less studied than those in higher dimensions. 
Although they are included in the analysis of \cite{hs}, 
\cite{hss} there is no canonical 
three-dimensional gauged supergravity (and hence dual conformal field theory) 
which has been studied. Ultimately this 
is because there are few known consistent truncations of 
spherical compactifications down to three dimensions which also admit 
anti-de Sitter metrics as solutions \cite{clp}.

For example, if one starts from the common bosonic sector of ten-dimensional 
supergravity theories and compactifies on a seven-sphere, it is consistent
to truncate to just the massless modes but the resulting gauged supergravity 
theory admits domain wall rather than anti-de Sitter solutions \cite{clp}. Other 
consistent spherical compactifications involving (for simplicity) only a 
small set of higher-dimensional fields are extremely rare. In fact, the only 
other example which has been discussed explicitly is the compactification of 
the six-dimensional bosonic string low-energy effective Lagrangian on a 
three-sphere to three dimensions \ci{clp}.

Working at the purely classical level we restrict to the simple Lagrangian
\be
\cL_{6} = R - \half (\del\phi)^2 - {\textstyle {1 \over{12}}} 
e^{-a \phi} (F_3)^2, \label{1a}
\ee
where $F_3$ is the Kalb-Ramond three-form and $a = \sqrt{2}$. The value of the 
constant $a$ is critical in ensuring that the spherical compactification can 
be truncated consistently just to massless modes \cite{clp}. 
Now the conformal anomaly term adds a term of the form
\be
\d \cL = - \half q^2 e^{\fr{1}{2} a \phi} 
\ee
to the Lagrangian. This extra contribution is a cosmological term;  
the spherical compactification remains consistent with this term. 
However if we include this term $AdS_3 \times S^3$ is not a solution of the
equations of motion; we have to switch on the six-dimensional dilaton and
consider domain wall solutions in three dimensions instead. Since here we are
working purely at the classical level, and do not need to appeal to any
microscopic description, there is no objection to omitting the conformal
anomaly term and working purely with (\ref{1a}).

The six-dimensional equations of motion are then
\bea
R_{mn} &=& \half \del_m \f \del_n \f + \qu e^{-\sqrt{2} \f} F_{mpq} 
F_{n}^{\hspace{1mm} pq} - {\textstyle {1\over{24}}} 
e^{-\sqrt{2} \f} F_{pqr} F^{pqr} g_{mn}; \nn \\
D_{m} \del^{m} \f 
&=& - {\textstyle {\frac{\sqrt{2}}{12}}} 
e^{-\sqrt{2} \f} F_{pqr} F^{pqr};  \\
d ( e^{-\sqrt{2} \f} \ast F_3) &=& 0. \nn
\eea
Here we work in Lorentzian signature and 
the index $m,n$ runs between $0$ and $5$. These equations manifestly 
admit a solution with constant dilaton and self-dual three-form such that 
\bea
ds_6^2 &=& ds_3^2 + l^{-2} d\W_3^2; \nn \\
F_3 &=& 2 l \epsilon_3 + 2 l^{-2} d\W_3,
\eea
where $d\W_3^2$ is the unit metric on the three-sphere and $ds_3^2$ is an 
Einstein metric of constant negative curvature satisfying 
$R_{\m\n} = -2 l^2 g_{\m\n}$. 

We use the following Kaluza-Klein ansatz for the spherical reduction
\bea
ds_6^2 &=& Y^{\fr{1}{4}} \left ( \D^{\fr{1}{2}} 
ds_3^2 + l^{-2}\D^{-\fr{1}{2}} T_{ij}^{-1} 
D\mu^i D \mu^j \right ); \nn \\
e^{\sqrt{2} \phi} &=& \D^{-1} Y^{\fr{1}{2}}; \la{eq4}\\
e^{-\sqrt{2} \phi} \ast F_3 &=& - l U \ep_3 + l^{-1} 
T_{ij}^{-1} \ast DT_{jk} \wedge 
(\m^k D \m^i) - \half l^{-2} 
T_{ik}^{-1} T_{jl}^{-1} \ast F^{ij} \wedge D \m^k \wedge
D \m^l \nn \\
&& + \textstyle{\frac{m}{6}} 
l^{-2}\ep_{ijkl} \m^i D \m^j \wedge D \m^k \wedge D \m^l
; \nn \\
F_3 &=& m Y \ep_3 + \textstyle{\frac{1}{6}} 
\ep_{ijkl} \left ( l^{-2} U \D^{-2} D \m^{i} 
\wedge D \m^{j} \wedge D \m^k \m^l - 3 l^{-1} \D^{-1} F^{ij} \wedge 
D \m^{k} T_{l i' } \m^{i'} \right .\nn \\
&& \left .- 3 l^{-2} \D^{-2} D\m^i \wedge D \m^j 
\wedge DT_{ki'} T_{l j'} \m^{i'} \m^{j'} \right ), \nn
\eea
with the following definitions
\bea
\m^i \m^i = 1; && \D = T_{ij} \m^i \m^j; \nn \\
U = 2 T_{ik} T_{jk} \m^i \m^j - \D T_{ii}; && Y = \det (T_{ij}); \nn \\
D \m^i &=& d\m^i + l A^{ij}\m^j; \la{eq5} \\
DT_{ij} &=& dT_{ij} + l A^{ik} T_{kj} + l A^{jk} T_{ik}; \nn \\
F^{ij} &=& dA^{ij} + l A^{ik} \wedge A^{kj}. \nn
\eea
In addition, $\ep_3$ in (\re{eq4}) is the volume form of the three-dimensional 
metric. The indices $i,j$ are $SO(4)$ indices and hence run between 
$1$ and $4$. The three-sphere is parametrised in this ansatz by a set of
direction cosines. The effective three-dimensional
Lagrangian is then
\bea
\cL_3 &=& R_3 - \textstyle{\frac{1}{16}} Y^{-2} (\del Y)^2 
- \qu \td{T}_{ij}^{-1} (D \td{T}_{jk}) \td{T}_{kl}^{-1} (D \td{T}_{li}) 
- \half m^2 Y \\
&& - \textstyle{\frac{1}{8}} Y^{-\fr{1}{2}} \td{T}_{ik}^{-1} 
\td{T}_{jl}^{-1} F^{ij} F^{kl} 
- \half l^2 Y^{\fr{1}{2}} (2 \td{T}_{ij}\td{T}_{ij} - \td{T}_{ii}^2), \nn
\eea
where we have found it convenient to introduce the unimodular $\td{T}$ where
\begin{equation}
T_{ij} = Y^{\fr{1}{4}} \td{T}_{ij}.
\end{equation}
Note that although
in toroidal compactifications we can dualise the three-dimensional 
gauge potentials to scalars this is not possible for the gauged theory. 
The resulting three-dimensional equations of motion are
\bea
R_{\m\n} &=& \textstyle{\frac{1}{16}} Y^{-2} (\del_{\m}Y)(\del_{\n} Y) 
+ \qu \td{T}_{ij}^{-1} (D_{\m} \td{T}_{jk}) \td{T}_{kl}^{-1} 
(D_{\n} \td{T}_{li}) + \half m^2 Y g_{\m\n} \nn \\
&& + \half l^2 Y^{\fr{1}{2}} 
(2 \td{T}_{ij} \td{T}_{ij} - \td{T}_{ii}^2)g_{\m\n} 
 + \qu Y^{-\fr{1}{2}} \td{T}_{ik}^{-1} \td{T}_{jl}^{-1} F^{ij}_{\m\r} 
F_{\n}^{kl\r};  \la{se1} \\
D_{\m} (Y^{-2} \del^{\m} Y) &=& - \half Y^{-\frac{3}{2}} 
\td{T}_{ik}^{-1} \td{T}_{jl}^{-1} F^{ij}_{\m\n} F^{kl\m\n} 
+ 2 l^2 Y^{-\fr{1}{2}} 
(2 \td{T}_{ij} \td{T}_{ij} - \td{T}_{ii}^2) + 4 m^2; \nn \\ 
D_{\m} (\td{T}_{ik}^{-1} D^{\m} \td{T}_{kj}) 
&=& 2 l^2 Y^{\fr{1}{2}} ( 2 \td{T}_{ik} \td{T}_{jk} - \td{T}_{ij} \td{T}_{kk}) 
- Y^{-\fr{1}{2}} \td{T}_{lm}^{-1} \td{T}_{ik}^{-1} (F_{\m\n}^{lk} F^{mj \m\n}) 
\nn \\
&& - \qu \d_{ij} \left (2l^2 Y^{\fr{1}{2}} (2 \td{T}_{lk} \td{T}_{lk} 
- (\td{T}_{kk})^2) 
- Y^{-\fr{1}{2}} \td{T}_{lm}^{-1} \td{T}_{nk}^{-1} F^{lk}_{\m\n} 
F^{mn\m\n} \right ); \nn \\
D( Y^{-\fr{1}{2}} \td{T}_{ik}^{-1} \td{T}_{jl}^{-1} \ast F^{kl}) &=& 
-2l \td{T}^{-1}_{k[i} \ast D \td{T}_{j]k} - \half m \ep_{ijkl} F^{kl}. \nn 
\eea
Here each derivative $D$ is of the appropriate fully covariantised form.
The (locally) $AdS_3$ solutions of these equations are recovered by setting 
\be
m^2 = 4 l^2; \hsp \td{T}_{ij} = \d_{ij}; \hsp Y = 1; \hsp F^{ij} = 0. \la{eq6}
\ee
We would like to determine the IR divergences of the action in both six and 
three dimensions. Let us consider first the three-dimensional case. To 
evaluate the on-shell divergences we need to solve the field equations in
the vicinity of the AdS boundary to sufficient order to determine all
IR divergences. To do this we use the by now well known results of Fefferman 
and Graham \ci{gl}, \ci{fg} 
that any $(d+1)$-dimensional metric of negative curvature admits an 
asymptotic expansion near the boundary of the form
\be
ds_{(d+1)}^2 = \frac{d\r^2}{l^2 \r^2} + \frac{1}{l^2 \r^{2}} 
g_{\a\b} dx^{\a} dx^{\b}, \la{eq7}
\ee
where the $d$-dimensional metric $g$ admits an expansion 
\be
g = g^0 + \r^2 g^2 + ... \r^d g^d + h^d \r^d \ln \r + ... \la{eq8}
\ee
The logarithmic term appears only when $d$ is even and only even powers of 
$\r$ appear up to this order. $l$ is the scale parameter which appeared above. 
We will also need to expand the matter fields about their background (or 
asymptotic) values given in (\re{eq6}) and then solve the full set of
coupled equations. This turns out to be much easier than one might have 
naively expected. The key simplification is that both the vectors and all the 
scalars have to admit expansions that start at powers of $\r$ too high to 
contribute to the field equations or to the action, to the order necessary 
to determine all divergences. The chain of arguments required to determine 
this is quite complicated and involves all the field equations. 

Firstly let us expand the two dimensional metric from (\ref{eq7}) as
\be
g_{\a\b} = g^{0}_{\a\b} + \r^2 g^2_{\a\b} + \r^2 \sum_{p=1}  
h^{p}_{\a\b} \ln \r + ... ,
\ee
where the ellipses denote terms of higher order where the expansion breaks 
down. Although for purely 
gravitational divergences we need only include the first logarithmic term
$h^{1}_{\a\b}$, it is known that with matter source terms one can have 
further logarithmic terms. Explicitly calculating the curvature of this metric
we find the following
\bea
R_{\r\r} &=& - \frac{2}{\r^2} - l^2 \left ( \tr((g^0)^{-1} h^1) 
- \tr ((g^0)^{-1} h^2) - 2 \ln \r \tr( (g^0)^{-1} h^2) \right .\nn \\
&& \left . - 3 \ln \r (1 + \ln \r) \tr( (g^0)^{-1} h^3) \right ) + ...;  \\
(g^0)^{\a\b} R_{\a\b} &=& - \fr{4}{\r^2} + l^2 R^0 - l^2 \left (
\tr((g^0)^{-1} h^2) - 3 \ln \r \tr ((g^{0})^{-1}h^3) \right )+ .... \nn 
\eea
where we have retained only terms up to order $h^3$ and $R^0$ is the curvature of the metric $g^0$. All contractions are taken in the metric $g^0$. 

Now using the Einstein equation in (\re{se1}) we can deduce that to
preserve the requisite form of the metric (\re{eq6}) the matter fields must
satisfy 
\bea
\d Y & = & \r^2 \sum_{p=0} Y^{(p)}(x^{\a}) (\ln\r)^p + .. \nn \\
\d\td{T}_{ij} & = & \r \sum_{p=0} t_{ij}^{(p)} (\ln \r)^p 
+ \r^2 \sum_{p=0} \t_{ij}^{(p)} (\ln \r)^p + ... ; \la{q1}\\
F^{ij}_{\m\n} F^{\m\n}_{ij} & \le & O(\r^2), \nn
\eea
where we will give a more detailed expansion of the vector fields later. 
The first two expansions are perturbations about the background values
given in (\re{eq6}). Since $\td{T}$ is unimodular, expanding out the 
determinant we find that each $t_{ij}^{(p)}$ must be traceless whilst 
\be
\t_{ii}^{(r)} = \half \sum_{p+q = r} t_{ij}^{(p)} t_{ji}^{(q)}.
\ee
From these constraints we find that 
\be
(2 \td{T}_{ij} \td{T}_{ij} - \td{T}_{ii}^2) = -8 + O(\r^3),
\ee
and it is this which, using the equation for 
$Y$, forces the perturbation in $Y$ to be at least as small as $\r^2$. 
This follows from writing the equation as 
\be
D_{\m}(Y^{-2} \del^{\m}Y) = 8 l^2 \d Y + O(\r^2),
\ee
from which it is easy to 
show that $\d Y$ cannot be of order $\r$. Given that $\d Y$ 
is of order $\r^2$ or smaller, it turns out that it 
will not contribute to the Einstein 
or other field equations to the required order and can henceforth be
neglected. 

A similar, though slightly more subtle, argument can now be used to show that
$t_{ij}^{(p)} = 0$. Expanding out the equation for $\td{T}_{ij}$, we find
that $t_{ij}^{(p)}$ must satisfy
\be
\sum_{p=0} D_{\r} D^{\r} ( \r t_{ij}^{(p)} (\ln \r)^p) \le O(\r^2). \la{eq9}
\ee
This can evidently not be satisfied for non-zero $t_{ij}$ when the gauge field
vanishes. It can also not be satisfied even when the gauge field is 
switched on. To demonstrate this, let us retain only the $t_{ij}^{0}$ term
for simplicity. Using the Einstein equations to constrain the gauge field 
strengths to be at least as small as in (\re{q1}) the gauge potentials 
should be expanded as
\bea
A^{ij}_{\r} &=& \r^{-1} \sum_{p=0} \b^{ij(p)}(x^{\a}) (\ln \r)^p; \la{gf}\\
A^{ij}_{\a} &=& \sum_{p=0} \a^{ij(p)}_{\a} (x^{\a}) (\ln \r)^p. \nn
\eea
Note that the $\a$ components must be of this order to ensure the
absence of $\r^{-2}$ terms in the field strength. 
Substituting the gauge potential into the equation (\re{eq9}), and for 
simplicity of the argument retaining only the $p=0$ terms in the potentials,
we find the following constraint
\bea
v^{ij} & \equiv & t^{ij} + \b^{ik} t^{kj} + \b^{jk} t^{ki}; \\
v^{ij} &=& \b^{ik} v^{kj} + \b^{jk} v^{ki}, \nn
\eea
where we have suppressed $p$ labels. $v^{ij}$ is a 
traceless symmetric matrix, like $t^{ij}$. These constraints can 
evidently not be satisfied for non-zero $v^{ij}$ or $t^{ij}$ since, for 
example, the second constraint implies 
\be
\det{\b} = \det{(1+\b)}.
\ee
It is straightforward to generalise these arguments to include non-zero 
values of $p$. Thus we have proved that the expansion of $\td{T}$ starts
at order $\rho^2$ and to the required order the three-dimensional field 
equations reduce to 
\bea
R_{\r\r} &=& - 2 l^2 \r^{-2} - \qu l^2 \r^2 (g^0)^{\a\b}
F^{ij}_{\r\a} F^{ij}_{\r\b} + ...; \nn
\\
(g^0)^{\a\b} R_{\a\b} &=& -2 l^2 \r^{-2} (g^{0})^{\a\b} g_{\a\b} 
- \qu l^2 \r^2 (g^0)^{\a\b} F^{ij}_{\r\a} F^{ij}_{\r\b} + ...; \la{se2} \\
D(\ast_3 F^{kl}) &=& - l \ep_{ijkl} F^{kl} + .... \nn
\eea
For the gauge fields we have used the fact that, following from (\ref{gf}),
the field strengths can be expanded as 
\bea
F^{ij}_{\a\b} &=& \sum_{p=0} f^{ij(p)}_{\a\b} (ln \r)^p + ...; \\
F^{ij}_{\r\a} &=& \r^{-1} \sum_{p=0} g^{ij(p)}_{\r\a} (\ln \r)^p + ... , 
\nn 
\eea
and hence the components $F^{ij}_{\a\b}$ will not contribute to the 
required order. 

Solving the last of the equations in (\re{se2}) would be complex but this
turns out to be unnecessary since the two Einstein equations force 
\bea
\tr ( (g^0)^{-1} h^{p}) &=&  0 \hsp \forall p; \nn \\
F^{ij}_{\a\r} F^{ij}_{\a\r} &<& O(\r^{-2}); \\
\tr( (g^{0})^{-1} g^2) &=& - \half l^2 R^0, \nn
\eea
so that, as claimed above, the only IR divergences of the action are the 
gravitational ones. It is important in what follows that we have excluded
$\r$ corrections to any of the fields.

\bigskip

The three-dimensional action is 
\be
\CS = \frac{1}{2 \k^2} \int \sqrt{g} \cL_3  - 
\frac{1}{\k^2} d^2x \int K \sqrt{h},
\ee
where as usual the second term is the Gibbons-Hawking boundary term with $K$
the trace of the second fundamental form of the boundary. Explicitly 
calculating the IR divergences by cutting off the boundary at 
$ \r = \ep$, we find
\be
\CS_{\rm{div}} =  - \frac{1}{\k^2} \int d^2x \sqrt{g^0} \left ( 
\frac{1}{l \ep^2} - (R^0 l ) \ln \ep \right ),
\ee
which leads to the following counterterm action 
\be
\CS_{\rm{cov}} =  \frac{1}{\k^2} \int d^2x \sqrt{h} \left ( l^{-1} 
- l R[h] \ln\ep \right ),
\ee
in terms of the induced metric on the boundary $h$ and its curvature $R[h]$.
This completes the calculation of divergences and counterterms in three
dimensions. Although the final answer was very simple, involving only the 
induced metric on the boundary and its curvature, we solved the full 
set of equations in order to find all possible divergences and explicitly 
showed that, for example, there can be no terms in odd powers of $\r$ in 
the metric. This is significant in what follows.

\bigskip

Now let us consider the divergences and counterterms from the higher 
dimensional perspective. At first sight it seems as though the uplift to
six dimensions will be trivial since all matter field perturbations are
subleading. However, although we can and will set  
the gauge fields and $Y$ perturbations to zero, we should retain the
perturbation  
\be
\td{T}_{ij} = \d_{ij} + \r^2 \t_{ij}(x^{\a}),
\ee
where $\t_{ij}$ is a traceless symmetric matrix. Although 
this perturbation does not 
contribute to the IR divergences it deforms the three sphere when we uplift
the solution to six dimensions and is the leading order field perturbation.

Using the Kaluza-Klein ansatz of (\ref{eq4}) we can now lift our 
three dimensional fields up to six dimensions. This gives
\bea
ds_6^2 &=& ( 1 + \half \r^2 G) ds_3^2 + 
l^{-2} (1 - \half \r^2 G) \left ( g^0_{ab} + \r^2 g^2_{ab} de^a de^b \right ) 
+ ... ; \nn \\
g^2_{ab} &=& - G g^0_{ab} - \half G_{;(ab)}; \nn \\
G_{;a}^{\hspace{2mm} a} &=&  - 8 G \hsp \rightarrow \hsp 
\tr( (g^0)^{-1} g^2) = G; \la{eq11} \\
e^{\sqrt{2} \f} &=& (1 - \r^2 G + ...) ; \nn \\
F_{3} &=& 2 l \ep_3 + 2 l^{-2} (1- 2 \r^2 G) \eta_3^{0} - 4 l^{-2} d(\r^2 G) 
\wedge \del\eta_{3}^{0}  ..., \nn
\eea
where in the first line $ds_3^2$ is the three-dimensional metric already 
determined and $g^0_{ab}$ is an Einstein metric on the unit three-sphere, 
with coordinates $x^a$ where $a=1,3$. The deformation of the spherical 
metric is defined in the second line, using the covariant
derivative on the three-sphere and a function $G$ which is related to the
traceless symmetric tensor $\t_{ij}$ already defined as
\be
G(x^{\b};x^{a}) = \t_{ij}(x^{\b}) \m^{i} \m^{j},
\ee
and is thus an $l=2$ harmonic on the three-sphere. This is the implication 
of the third line of (\re{eq11}). In the last line, 
$\ep_3$ is the volume form of the full three-dimensional metric whilst 
$\eta_3^{0}$ is the volume form of the unit three-sphere. $\del\eta_{3}^{0}$
is the interior derivative of the volume form; this final term in $F_3$ 
ensures that it is closed. 

With this expansion of the six-dimensional fields we can proceed 
to calculating the divergences and counterterms. We could 
of course have worked out the expansions of the fields by using an 
appropriate ansatz to solve the six-dimensional equations directly. 
There are a number of reasons for not doing this. 
Firstly, it is quite difficult to find an appropriate ansatz that is
sufficiently general. Secondly, even given an appropriate ansatz, the 
six-dimensional field equations are actually slightly more involved than
the dimensional reduced equations, essentially because of the explicit 
spherical dependence. Thirdly, we do of course already know the 
three-dimensional
equations explicitly whereas for any six-dimensional ansatz we would need
to work them out! 

Finally, and most importantly, the explicit frame
dependence of the six-dimensional counterterm action and its 
relationship to the covariant counterterm action in three dimensions is
much clearer when we follow the route of uplifting from three dimensions. 
Lack of covariance is the crucial issue and it is important to identify
clearly the origins of the problem. 

The observant reader would be right to have misgivings at this stage, 
though, for the following reason. Although all solutions of the
three-dimensional field equations are, by construction, solutions of
the six-dimensional equations, the reserve is certainly not true. Any
six-dimensional solution which involves three-dimensional fields not
included in the supergravity multiplet (i.e. massive Kaluza-Klein excitations)
will not solve the three-dimensional field
equations used here and will hence be excluded from our analysis. We
are not ignoring this point and will shortly return to discuss it in detail. 

For now let us proceed with the uplift of the three-dimensional solutions. 
One would expect that the six-dimensional action should be taken to be, 
as usual,
\be
\CS_{6} = \frac{1}{2 \k_6^2} \int d^6 x \sqrt{g} \cL_{6} 
- \frac{1}{\k_6^2} \int d^5x K \sqrt{h},
\ee
where we have used the six-dimensional coupling constant and included the
Gibbons-Hawking boundary term. However, although this action 
can be shown to
reproduce the same field equations as the three-dimensional action when 
one substitutes the Kaluza-Klein ansatz, it does not reproduce the 
same action. The key to the discrepancy is the bulk term: the on-shell
value is
\be
\CS_{\rm{bulk}} = -\frac{1}{24 \k_6^2} \int d^6 x \sqrt{g} 
e^{-\sqrt{2} \f} (F_{3})^2.
\ee
However, to leading order $F_3$ is self-dual and the total contribution of 
the bulk term to the divergences is
\be
\CS_{\rm{bulk}} = - \frac{1}{\k_6^2 l^4} \ln \ep \int d^2x \sqrt{g^0} 
\int d^3x \sqrt{\eta^0} G(x^{\a};x^{a}) = 0,
\ee
where $\eta^0$ is the measure on the three-sphere. Since $G$ is an
$l=2$ harmonic, the integral of $G$ over the three-sphere vanishes and 
hence there are no divergent contributions from the bulk term. The 
divergence from the Gibbons-Hawking boundary term is 
\be
\CS_{\rm{GH}} = - \frac{2}{\k_{6}^2 l^4 \ep^2} \int d^2 x \sqrt{g^0} 
\int d^3x \sqrt{\eta^0} = - \frac{4 \pi^2}{\k_6^2 l^4 \ep^2} \int d^2 x 
\sqrt{g^0}.
\ee
Thus the divergent part of the action does not agree with the 
three-dimensional action! There is no reason why it should since so far
we have demanded
only that the field equations are equivalent for a consistent reduction, 
not that the actions are
equivalent. Nonetheless there are other boundary terms that one could add 
to the six-dimensional action (or indeed the three-dimensional action) without
affecting the equations of motion. The term that is relevant here is 
\be
\d\CS = \frac{1}{12 \k_6^2} \int d\Sigma^{\s} A^{\m\n} F_{\m\n\s} 
e^{-\sqrt{2}\f}. \la{f1}
\ee
Note that although this term lives on the boundary it is {\it not} 
a counterterm
because it does not just involve quantities intrinsic to the boundary. It
depends on the embedding of the boundary hypersurface into the bulk. 

There is no reason {\it a priori} why we need to demand equivalence between 
higher and lower dimensional actions. We could proceed with renormalising
the higher dimensional action with no such boundary term: the renormalised 
action would differ from that for the same solution evaluated with three
dimensional fields, but the renormalisation procedure would of course 
still be consistent. We take the view here that it is more convenient
for the higher and lower dimensional actions to be equivalent.

\bigskip

The physical interpretation of adding such a term to the action is that we are
shifting between different thermodynamic ensembles, from one where the total
electric charge is fixed to one where it becomes a thermodynamic variable. The
addition of such a term is well-known in the context of, for example,
calculating the free energy for black hole metrics \ci{hr}. Let us 
illustrate this
in the best known case, namely four-dimensional Reissner-Nordstrom
black holes. It is straightforward to show by explicit calculation that the
action given by
\be
\CS = \frac{1}{2 \k_4^2} \int d^4 x \left ( R - \frac{1}{4}F^2 \right ) - 
\frac{1}{\k_{4}^2} \int d^3x K \sqrt{h}, \la{q2}
\ee
when evaluated on-shell for Reissner-Nordstrom electric black holes (using 
the usual
background subtraction methods) is equal to the free energy in a grand
canonical ensemble such that 
\be
\CS = - \beta F = \beta M - \beta Q \Phi_{h} - S_{h},
\ee
where $F$ is the free energy, $\beta$ is the inverse temperature, $M$ is the 
mass, $Q$ is the electric charge, $\Phi_{h}$ is the difference between the
electric potential at the horizon and infinity and $S_{h}$ is the entropy 
(all in appropriate units). If we wish to calculate the free energy in a
canonical ensemble in which the electric charge is fixed, then we have
to add an additional term to the action (\ref{q2}) of the form
\be
\d \CS = \frac{1}{4 \k_{4}^2} \int d\Sigma^{\s} A^{\m} F_{\s\m},
\ee
so that
\be
\CS + \d \CS =  \beta M - S_{h}.
\ee
It turns out that to ensure the divergent part of the 
on-shell action is the same in six dimensions as in three we need to work
in the canonical ensemble where the electric charge is fixed. This is an
interesting point which had not been noticed before and may be relevant in
considering stability of anti-de Sitter black holes from a higher-dimensional
perspective, where the choice of ensemble is quite subtle \ci{cg1}, \ci{cg2}. 

One could in principle show, using the Kaluza-Klein ansatz,
that the six-dimensional action with this boundary term reproduces the 
three-dimensional action. This would require us showing that the higher
and lower dimensional sets of Einstein equations are equivalent, which was
never explicitly proved for this reduction \ci{clp}. We 
checked this equivalence for a truncated set of three-dimensional fields,
the metric and a single active scalar, and the corresponding uplift; this
is an adequate check provided that, as expected, the Einstein equations
are also equivalent.  

With the addition of the boundary term, the total divergences of the six 
dimensional action are
\be
\CS_{\rm{div}} = - \frac{1}{\k_6^2 l^4} \int d^2 x \sqrt{g^0} \int d^3 x 
\sqrt{\eta^0} \left ( \ep^{-2} - l^2 R^0 \ln \ep \right ),
\ee
which evidently agrees with the three-dimensional divergences since
$\k_6^2 = 2 \pi^2 l^{-3} \k^2$. The logarithmic term originates in the form
of the two-form potential expanded near the boundary
\be
A_{\a\b} = - l^{-2} \r^{-2} - \half R^{0} \ln \r + ...
\ee
where we have chosen the gauge such that $A_{\r\a} =0$. 

To render the action finite we would now like to define an appropriate 
counterterm action to cancel the divergences. The counterterm action 
must not affect the equations of motion and must hence be defined entirely
in terms of quantities intrinsic to the regularising boundary. Furthermore,
we would hope that the action can be written in terms of covariant 
quantities on the boundary, such as the curvature of the induced metric.
Here however we run into a fundamental problem: the boundary is degenerate 
since the three sphere remains of finite size as we take the IR cutoff to
infinity. The implication of this is that although we can always subtract
off the divergences by defining an appropriate counterterm action this action
cannot in general be written in terms of covariant quantities of the
boundary fields. Since the boundary is degenerate there is no concept of
five-dimensional covariance.

\bigskip

Holography for degenerate boundary metrics was discussed in \cite{t} and we
will now review some of the arguments that appeared there. Let us first
consider an Einstein metric of negative curvature that admits an expansion 
of the Fefferman-Graham type (\re{eq7}) and (\re{eq8}). Then the induced
metric on the boundary is 
\be
h_{\a\b} = l^{-2} \r^{-2} g_{\a\b},
\ee
where $g$ is defined in (\re{eq8}). The divergences of the action will be
expressible as a power series
\be
\CS_{\rm{div}} = \int d^dx \sqrt{g^0} \left ( a_{d}(x^{\a}) \ep^{-d} + ... 
A(x^{\a}) \ln\ep + ... \right ). \la{q5}
\ee
A covariant counterterm action can be constructed using the following 
procedure. Each time we take a derivative in the boundary
we add a power of $\r$. Thus the 
measure is of order $\r^{-d}$, the Ricci curvature is of order $\r^2$, the
Ricci tensor squared is of order $\r^4$ and so on. To cancel the divergences
we can therefore construct an action of the form
\be
\CS_{\rm{ct}} = \int d^dx \sqrt{h} \left ( \a_{0} + \a_{2} R[h] 
+ \a_{4} ( R[h]^2 + \b R_{\a\b}[h]R^{\a\b}[h] ) + ... \right ), \la{q3}
\ee
where we have omitted logarithmic terms and the $\a_{i}$ are constants. Then
the first term cancels the $\ep^{-d}$ divergence whilst the term in $R[h]$ 
is needed to cancel the $\ep^{-d+2}$ divergence and so on. 

This procedure breaks down, however, for degenerate metrics. Suppose that the
boundary metric is of the form considered here, namely,
\be
h_{ef} = \pmatrix { l^{-2} \ep^{-2} (g^0_{\a\b} + \ep^2 g^2_{\a\b} + ..) 
& 0 \cr                  
0 & l^{-2} (g^0_{ab} + \ep^2 g^2_{ab} + ...) } \,
\ee
so that the measure is of order $\ep^{-2}$. The Ricci curvature of this
metric is 
\be
R_{ef} = \pmatrix { R^{0}_{\a\b} & 0 \cr
                  0 & (2 g^0_{ab} + \ep^2 R^2_{ab} + ...) } \,
\ee
where $R^{2}_{ab}$ is the first correction to the 
intrinsic curvature of the metric $g^0_{ab} + \ep^2 g^2_{ab}$. Its 
explicit form is 
\be
R^2_{ab} = \half G_{;ab} + \qu G g^0_{ab}.
\ee
The Ricci curvature of the boundary metric is 
\be
R[h] = 6 l^{2} + l^2 \ep^2 R^{0} + 2 l^2 \ep^2 G + ... \la{q4}
\ee
Since the metric is degenerate {\it all} curvature invariants start at 
leading order. One way to construct the action would be to repeat the 
usual procedure and take a counterterms of the form
\be
\CS_{\rm{ct}} = \int d^5x \sqrt{h} \left (a + b R[h] + c R[h]^2 + 
d R_{ef}[h]R^{ef}[h] + ... \right)
\ee
where we choose the coefficients to cancel divergences. The key difference 
from (\re{q3}) is that all terms will contribute to the cancel of the 
divergences at each order in $\ep$ and there is no ``natural'' choice
of coefficients. 

Perhaps
a more natural way of removing the divergences would be to write
the counterterm action in terms of not only the induced metric $h$ but also
the induced matter fields on the boundary. This is not necessary in this case 
but if there were other 
divergences arising from the bulk matter fields which could only be cancelled 
by including boundary matter fields in the counterterms it would become so. 
The induced matter fields on the boundary are the scalar field $\f^b$ and
potentials $B_{ef}$ and $B_{e}$ related to the bulk two-form potential as
\be
A_{mn} = B_{mn} + B_{[m} n_{n]}, \la{f2}
\ee
where $n$ is the unit normal to the boundary. With our previous gauge choice, 
the only relevant induced field is the two-form which satisfies
\bea
B_{\a\b} &=& - l^{-2}\ep^{-2} + \half R^{0} \ln \ep + ... \hsp \rightarrow 
\hsp H_{\a\b e} = 0; \nn \\
H &=& 2 l^{-2} \eta_{3}^{0} + O(\ep^3) .... ,
\eea
where $H$ is the field strength of $B$. From this field strength we can 
construct the covariant object
\be
H_{abc} H^{abc} = 24 l^2 ( 1 - G \r^2).
\ee
Putting this together with (\re{q5}) and (\ref{q4}) 
we find that we can express the counterterm action as 
\be
\CS_{\rm{ct}} =  \frac{l}{\k_6^2 } \int d^5x \sqrt{h} \left ( 1 - ( R[h] 
- \qu H^2) \ln \ep \right ). \la{q6}
\ee
So it seems that we have 
effectively evaded the potential problems caused by the metric being 
degenerate in this case: we have written the counterterms in 
terms of seemingly covariant quantities on the boundary. 
However although we have used ``covariant'' quantities
the action is not covariant: specific coordinate choices for the boundary
are still necessary for it to be applicable as we will see.

\bigskip

Given a renormalised action we need to clarify the conditions under
which this will indeed give a finite answer for solutions of the six
dimensional field equations because hidden in our analysis are various
assumptions, both obvious and subtle.

It is apparent that we cannot take an arbitrary six-dimensional
solution and expect (\ref{q6}) to render the action finite. This manifestly 
will not work for the six-dimensional Schwarzschild solution, for example,
even though this does 
satisfy the field equations. The counterterms will only work for
solutions where the metric is asymptotic to $AdS_3 \times S^3$. The 
subtleties arise in defining whether a spacetime does need indeed asymptote to
this. 

There are several equivalent ways of stating the conditions under 
which a metric is asymptotically $AdS$. If one can find a coordinate system
near the boundary such that the metric admits an expansion of the form
(\re{eq7}), (\re{eq8}) then the spacetime must be asymptotically AdS. Such
a condition is effectively a generalisation of the conditions on metric
components given in  in \cite{ht} for four dimensions and
in \cite{bh} for three dimensions
when $g^0$ is flat. One could also 
state the asymptotic conditions in terms of Penrose's definitions of
conformal infinity \ci{p}: if a spacetime of negative curvature has a regular
conformal boundary then the spacetime is asymptotically AdS. 

Let us try to define under what circumstances a solution of a particular
set of six-dimensional field equations is asymptotically $AdS_3 \times S^3$.
For the particular Lagrangian under consideration here we have effectively
derived the radial dependence of metric components such that the metric
is asymptotically of this form in a natural extension of the work of \cite{ht}
and \ci{bh}.  

There is however an important and subtle caveat, related to the lack
of covariance of a degenerate boundary. The divergences are covariant under
coordinate transformations of the $AdS_3$ and the $S^3$ parts of the
metric separately but are not covariant under the most general
coordinate transformations which mix the two parts of the metric. 

\bigskip

The simplest (and most physical) example of this is the following. In three
dimensions we have a scalar matrix $\td{T}_{ij}$ which lies in the symmetric
traceless representation of $SO(4)$ and corresponds to switching on $l=2$
spherical harmonics on the three-sphere. From the six-dimensional perspective
we can also switch on $l=1$ harmonics on the three-sphere: this would 
correspond in three dimensions to a vector representation $J_i$ of $SO(4)$.
Now $J_i$ is not part of the consistent truncation in three dimensions 
to supergravity \ci{clp} and is thus not included in our analysis. 
From the six-dimensional perspective, however, there is no reason not
to switch on $J_i$. This corresponds to having a non-zero
electric dipole moment of the $F_3$ charge distribution. 

Let us clarify here what we mean by dipoles in
this context. As usual the electric monopole moment of the charge
distribution is given by the (coordinate invariant) expression
\be
q = \int_{\Sigma_3} \ast F_3,
\ee
where $\Sigma_3$ is some appropriate closed 3-cycle. Higher pole moments
of the charge distribution can be defined once we have fixed a coordinate
choice. For a static charge distribution the 3-form will usually be defined
in terms of a function which admits an expansion of the form
\be
\phi = q_{0} \rho^2 + q_{1} g_1(x^a) \rho^3 + q_2 g_2 (x^a) \rho^4 + ... 
\la{pot1}
\ee
where $\rho$ has a first order zero on the conformal boundary and $g_{i}(x^a)$
are $i$-th harmonics on the sphere. Then $q_{i}$ can, with suitable 
normalisation, be identified as the $i$-pole moment of the charge
distribution, in direct analogy with classical electromagnetism.

Since the dipole moment is a coordinate dependent quantity it will always
be possible to eliminate it by making an appropriate coordinate
redefinition. Coordinate transformations can be used to eliminate all
the odd powers of $\rho$ in (\re{pot1}); this effectively removes $J_i$ and
brings us back
into the coordinate system of Fefferman and Graham. Since the 
counterterms are not covariant we should not however 
exclude the possibility of new divergences/counterterms being needed 
when we are not in the special coordinate system where these moments 
vanish. If we are in the dipole coordinate frame, and regulate the boundary
with the ``natural'' radial parameter, the renormalised action may no longer
be finite.

Put another way, we have determined all possible divergences from 
the three-dimensional perspective. However when we uplift to six dimensions 
and do a coordinate transformation which, for example,
induces a dipole moment new divergences may appear. 
In three dimensions the counterterm action
renormalises the action for {\it all} solutions of the field
equations. In six dimensions it does not because not all six-dimensional 
solutions solve the three-dimensional equations of motion: in general
we will need to switch on massive excitations in three dimensions as well. 

To show that there are such divergences in six dimensions let us
consider an explicit solution which has a dipole moment, a distribution of
dyonic black strings such that 
\bea
ds_6^2 &=& f^{-1} (-dt^2 + dx^2) + f (dr^2 + r^2 d\Omega_3^2); \nn \\
F_{3} &=& df^{-1} \wedge dt \wedge dx + \ast (df^{-1} \wedge dt \wedge dx), 
\la{st}
\eea
where $f$ is a harmonic function in flat space. An appropriate choice to
ensure that the metric is asymptotically $AdS_3 \times S^3$ with the same
curvature radius as before and a non-zero dipole moment is to take
\be
f = \frac{1}{l^2 r^2} ( 1 + g(x^a) r^{-1}),
\ee
where $g$ is an $l=1$ harmonic on the three-sphere and hence satisfies
$g_{;a}^{\hspace{2mm} a} = -3 g$. The spacetime is manifestly singular at
$r=0$ but this singularity does not affect our discussion of IR divergences.
(It will of course be relevant for UV divergences.)
The dipole moment of the charge
distribution can manifestly be removed by a coordinate redefinition 
\bea
\td{r} &=& r - \half g(x^a) + ..; \nn \\
d\td{\Omega}_3^2 &=& \left ( 1 + g(x^a) r^{-1} + ... \right ) d\Omega_3^2,
\eea
but the action divergences will not necessarily be preserved under such a
coordinate transformation. That is, if we regulate the boundary at some
fixed $\td{r}=\td{R}$, the renormalised action is guaranteed by our 
construction to be finite. However, if we regulate the boundary at some
fixed $r = R$ it is no longer guaranteed to be finite. 
Note also that in removing the dipole moment
we will in general induce a non-zero quadrupole moment; in other words
we will not be able to eliminate the $r^2$ corrections to the metric.

The bulk part of the action is exactly zero in this case since
the three form is self dual but the IR divergences from the boundary terms
when we cut off at $r=R \gg 1$ are
\be
\CS_{\rm{div}} = - \frac{1}{\k_6^2} \int d^2x \int d^3x \sqrt{\eta^0} 
\left (R^2 - {\textstyle{3 \over 2}} l^{-1} g(x^a) R + .... \right ),
\ee
where ellipses denote
finite terms. There is indeed an additional divergence linear in $R$ 
{\it but} this vanishes since the integral of any harmonic over the sphere 
vanishes. Evidently the next order terms will not be zero automatically
since the integral of, for example, $g^2$ over the sphere does not vanish 
but such terms are finite in $R$. There are in principle contributions to 
the action from the inner boundary cutting off the singularity, but these
are zero independently of how we cut off the singularity and so have no 
role here. 

Now we need to evaluate the counterterm action on this solution. For the
same reasons as above, the non-logarithmic counterterm does not have
any divergent contributions dependent on $g$ and is just given by
\be
\CS_{\rm{ct}} = \frac{1}{\k_6^2} \int d^2x \int d^3x \sqrt{\eta^0} 
\left (R^2 + ... \right).
\ee
However the logarithmic counterterm gives the following
\be
\CS_{\rm{ct}} = \frac{1}{l^2 \k_{6}^2} \int d^2x \int d^3x \sqrt{\eta^0} 
\left ( \textstyle{105 \over 2} g(x^a)^2 + \half (\del g)^2 \right ) \ln R ,
\ee
which does not vanish or cancel any other logarithmic
divergence. We cannot just discard the logarithmic term, since
it is necessary to cancel divergences for other solutions, 
so to cancel this divergence
we would need to subtract another logarithmic term from the counterterm action.
We have just shown explicitly that the counterterms
are not covariant in six dimensions, even though the counterterms are written 
in terms of apparently covariant quantities in the boundary. 

Now the question is: would it be possible to write down a (covariant) 
counterterm action which eliminates the divergences both for uplifted
solutions from three dimensions {\it and} six-dimensional solutions 
which have a dipole moment. After all, as we pointed out earlier, the 
choice of counterterms was not unique. We chose to cancel logarithmic
divergences with $R$ and $H^2$ terms but we could have used, for example,
$R$ and $R^2$ instead. 

It is in fact possible to fix the counterterms so that divergences are 
eliminated both when there is a dipole moment and when there is not. This
however is beside the point since there are still an {\it infinite} number of 
coordinate transformations of the boundary
that one could make in six dimensions for which
the resulting counterterms would not remove divergences. For example, starting
with pure $AdS_3 \times S^3$ in the Poincar\'{e} coordinate system, we could
do a coordinate transformation of the form
\be
\td{\r} = \r + a(x^a)_1 \r^2 +  a(x^2)_2 \r^3...,
\ee
where $a(x^a)_i$ are arbitrary functions on the sphere. Just plugging
this metric into the action, and regulating the boundary at some small fixed
$\td{\r}$ we will find new uncancelled divergences. We can adjust the
counterterms so that the divergences cancel for some choices of $a(x^a)_i$ 
but they will never cancel for all choices. However hard we
try we cannot get a covariant counterterm action. 

It is useful to understand this from two slightly different but complimentary
perspectives. Above we have been talking about bulk coordinate redefinitions:
we change the bulk coordinates and then regulate the boundary with the new 
``natural'' radial parameter.
However, one could keep the bulk solution in the same coordinate
system and just deform the boundary. The two perspectives are equivalent. 
From what we have said it is obvious
that under such a boundary deformation the counterterm action will not in
general still renormalise the action. To convince oneself of this, use an
explicit solution, such as
pure $AdS_3 \times S^3$ again, and cutoff the boundary in a non-standard way
with some function $f(x^{\a},\r,x^{a}) \rightarrow 0$; there must 
in general be uncancelled divergences when we explicitly evaluate our
``renormalised'' action.  

\bigskip

So given a generic metric which is asymptotically $AdS_3 \times S^3$ we
cannot renormalise the action using (\re{q6}) unless this solution is in the 
same class as the metric used to calculate (\re{q6}). One way to define this
is to demand that the metric admits a Kaluza-Klein reduction to three 
dimensions, using the ansatz (\re{eq4}), and explicitly identify
the three-dimensional fields, order by order in $\r$. 
It is easy to convince oneself
that (\re{st}) does not admit such a reduction until one removes the
dipole moment by a coordinate transformation.

In not every case will a coordinate redefinition be necessary to eliminate
divergences. Where the metric corrections arise from gauge transformations
or diffeomorphisms 
of three dimensional fields, the corrections are innocuous and will certainly
not lead to additional divergences in the action. 
An explicit example is the following. Consider the six-dimensional metric
\bea
ds_6^2 &=& \frac{l^2 r^2}{F} \left ( -(1 - \frac{2mF}{r^2})dt^2 
+ dy^2 \right ) + \frac{r^2}{l^2 \left ( (r^2 + a^2)(r^2 + b^2) - 2 m r^2 
\right )} dr^2 \nn \\
&& - 2 \left ( b \cos^2 \theta d\psi + a \sin^2 \theta d\phi \right ) dt 
- 2 \left (a \cos^2 \theta d\psi + b \sin^2 \theta d\phi \right ) dy \\
&& l^{-2} \left ( d\theta^2 + \sin^2 \theta d\phi^2 + \cos^2 \theta d\psi^2 
\right ); \nn \\
F^{-1} &=& \left ( 1 + \frac{a^2 \cos^2 \theta}{r^2} 
+ \frac{b^2 \sin^2 \theta}{r^2} \right ). \nn
\eea
This is a metric for (the near horizon geometry of)
rotating dyonic strings and is a solution to the field equations, with
appropriate choice of three-form. To use the counterterm action (\ref{q6}) we
first need to eliminate the crossterms to sufficient order which in this
case can be achieved simply by shifting the angular variables \ci{cl}
\bea
d\td{\psi} &=& d\psi - l^2 \left (b dt + a dy \right ); \\
d\td{\phi} &=& d\phi - l^2 \left (a dt + b dy \right ), \nn 
\eea
so that the metric becomes
\bea
ds_6^2 &=& - 
\frac{l^2 \left ( (r^2 + a^2)(r^2 + b^2) - 2 m r^2 \right )}{r^2} dt^2 
+ l^2 r^2 ( dy - \frac{ab}{r^2} dt)^2 \\
&& + \frac{r^2}{ l^2 \left ( (r^2 + a^2)(r^2 + b^2) - 2 m r^2 \right )} dr^2 
+ l^{-2} \left ( d\theta^2 + \sin^2 \theta d\td{\phi}^2 
+ \cos^2 \theta d\td{\psi}^2 \right ), \nn
\eea
in which form it is apparent that the geometry is a direct product of
two three-dimensional spaces. The coordinate transformation amounts to a
trivial gauge transformation of Abelian gauge fields in three dimensions. 

All of the above has focussed on the explicit metric dependence of the 
counterterm action. 
This is not the only way in which the six-dimensional action is not covariant:
it manifestly 
also depends on the explicit gauge choice for the two-form potential. 
This follows from (\re{f1}), (\re{f2}) and (\re{q6}): one can find gauge
transformations of the bulk two-form potential 
which do not leave the divergent part of the action invariant. Again this
is related to the lack of full six-dimensional covariance. So not only do
we have to bring the metric into the prescribed form but also we have to bring
the two form potential into the same gauge as chosen above. 

The conclusion of our analysis is the following. Given an asymptotically
$AdS_3 \times S^3$ solution of the six-dimensional field equations we 
can renormalise the action provided that the coordinate frame is such that
all fields can be matched directly to the perturbative expansions of
the three-dimensional fields given here and we regulate the boundary suitably.
The most obvious regulation is $\rho = \ep$ but other regulations are allowed 
provided that the ``crossterms'' between $AdS_3$ and $S^3$ coordinates
are sufficiently subleading. Three-dimensional covariance means that we
can certainly also regulate with any function of $AdS_3$ coordinates which has
a first order zero on the (degenerate) conformal boundary.
 
If the bulk fields cannot be directly 
matched we should change coordinates before evaluating the counterterms,
and again cutoff the boundary suitably. Equivalently we could cutoff
the boundary not with the natural radial parameter but in such a way as to
be equivalent to a regulated boundary in our preferred coordinate frame.
In the next section we will give an 
explicit demonstration of how this process works. 
Thus although we would have hoped to express everything covariantly from
a six-dimensional perspective we are forced back to three dimensions
if we want covariance of the counterterm action. Explicit calculations are 
more naturally carried out in the lower dimension, although the higher 
dimensional procedure can be made consistent and well-defined.

\section{Divergences and counterterms for $AdS_5 \times S^5$ spacetimes}

Of greater interest than the toy example discussed in the previous section
are the compactifications of eleven dimensional supergravity on $S^4$ and
$S^7$ and of type IIB supergravity on $S^5$. In each case one can truncate
the higher dimensional Lagrangian to the metric and $p$-form and show that
there is a consistent spherical reduction retaining only the massless 
modes \ci{clpst}, \ci{nvv}.
So we could proceed as before, solving for all the lower dimensional 
divergences (which will not just be purely gravitational) and then uplifting
to derive the higher dimensional counterterms. However the issues and analysis
would simply repeat the previous section: we will have the same lack of 
covariance and be forced to choose a particular gauge for the $p$-form
and a particular coordinate frame (and/or regulation) 
in order to apply our higher-dimensional counterterm action. 

In this section we will focus instead on answering a question that motivates
trying to renormalise the action directly in higher dimensions: given
an arbitrary distribution of D3-branes
\bea
ds_{10}^2 &=& \frac{1}{\sqrt{D}} dx \cdot dx_4  + \sqrt{D} dy 
\cdot dy_6; \nn \\
F_5 &=& dD^{-1} \wedge dx_4 + (\ast_6 \hspace{2mm} dD), 
\la{d1}
\eea
where $D$ is an harmonic function on $R^6$, can we calculate the 
renormalised action? This D3-brane distribution is a solution of the equations
of motion for type IIB supergravity truncated to the metric and self-dual
5-form $F_{5}$ with action
\be
\CS_{10} = \frac{1}{2\k_{10}^2} \int d^{10}x \sqrt{g} \left (R 
- \frac{1}{4 \cdot 5!} F_{5}^2 \right ) 
- \frac{1}{\k_{10}^2} \int d^9x K \sqrt{h},
\la{k1}
\ee
where as usual we impose by hand the self-duality condition. As before to
reproduce the lower dimensional action upon Kaluza-Klein reduction we need to
work in the canonical ensemble and add a boundary term
\be
\d\CS_{10} = \frac{1}{8 \cdot 5! \k_{10}^2} 
\int d\Sigma^{\s} A^{\m\n\r\t}F_{\s\m\n\r\t}. \la{k2}
\ee
From supersymmetry the action should be zero for any distribution
of positive tension branes since the renormalised action is zero when
all the branes are located at the same point and separating the branes
preserves supersymmetry. The goal of this section is to prove this.

Ten-dimensional fields can be reduced
to five dimensions using a Kaluza-Klein ansatz analogous to (\re{eq4}):
\bea
ds_{10}^2 &=& \D^{\frac{1}{2}} ds_5^2 + l^{-2} \D^{-\frac{1}{2}} 
T_{ij}^{-1} d\m^{i} d\m^{j}; \nn \\
F_5 &=& G_5 + \ast G_5; \la{p1} \\
G_5 &=& - l U \ep_5 + l^{-1} (T_{ij}^{-1} \ast d T_{jk}) \wedge 
(\m^{k} D \m^{i}); \nn \\
\ast G_{5} &=& \frac{1}{5!} \ep_{i_1..i_6} \left ( l^{-4} U \D^{-2} 
D \m^{i_1} \wedge ... \wedge D \m^{i_5} \m^{i_{6}} \right . \nn \\
&& \left . - 5 l^{-4} \D^{-2} D \m^{i_1} \wedge .... \wedge D\m^{i_4} \wedge 
DT_{i_5j} T_{i_6k} \m^j \m^k \right ), \nn 
\eea
where as in the previous section
\be
U = 2 T_{ij} T_{jk} \m^i \m^k - \D T_{ii}, \hsp \D = T_{ij} \m^i \m^j, 
\hsp \m^i \m^i = 1.
\ee
Here the indices $i,j$ are $SO(6)$ indices and $T_{ij}$ is a symmetric
unimodular tensor. We have truncated the $SO(6)$ Yang-Mills gauge
fields to zero; this is consistent with the field equations. Truncating the
vectors is permissible since the analysis will never be covariant from
the ten-dimensional perspective; we will always 
have to choose a particular coordinate
frame. Truncating the vectors simply restricts this frame to be one in
which all cross-terms vanish and restricts further 
the class of ten-dimensional solutions to which we can apply the 
renormalised action.

The equations of motion can be derived from the five-dimensional Lagrangian
\be
\cL_{5} = R - \qu T_{ij}^{-1} (\del T_{jk}) T_{kl}^{-1} (\del T_{li}) 
- \half l^2 \left (2 T_{ij} T_{ij} - (T_{ii})^2 \right ).
\ee
Expanding the fields in the usual way and solving the equations of motion
we find that 
\bea
T_{ij} &=& \d_{ij} + \r^2 t_{ij} + \r^2 \ln \r \td{t}_{ij} + \r^4 T_{ij}^{0} +
\r^4 \ln \r T_{ij}^{(1)} + \r^4 (\ln \r)^2 T_{ij}^{(2)}; \nn \\
T_{ii}^{(0)} &=& \half t_{ij}t_{ji}, \hsp 
T_{ii}^{(1)} = t_{ij} \td{t}_{ji}, \hsp
T_{ii}^{(2)} = \half \td{t}_{ij} \td{t}_{ji}; \nn \\
\tr( (g^0)^{-1} g^4) &=& - {\textstyle{1 \over 48}} \td{t}_{ij} \td{t}_{ji}
- {\textstyle{1 \over 6}} t_{ij} t_{ji}; \la{x1} \\
\tr( (g^0)^{-1} h^{(1)}) &=& - {\textstyle{1 \over 3}} t_{ij} \td{t}_{ji}; \hsp
\tr( (g^0)^{-1} h^{(2)}) = - {\textstyle{1 \over 6}} 
\td{t}_{ij} \td{t}_{ji}, \nn 
\eea
where we have assumed $g^0$ is flat since this is the case of interest here.  
Here $t_{ij}$ corresponds to a vacuum expectation value for a
dual scalar operator of dimension two whilst $\td{t}_{ij}$ 
corresponds to a source for such an operator.  

This is all we need to determine the
action divergences, although the field equations do fix the $T^{(i)}_{ij}$ 
entirely (although as usual only the trace of $g^4$ is determined \ci{hss}). 
Explicitly calculating the IR divergences in the action we find
\be
\CS_{\rm{div}} = - \frac{1}{l^3 \k^2_5} \int d^4x \sqrt{g^0} 
\left ( 3 \ep^{-4} + {\textstyle{1 \over 8}} 
\td{t}_{ij}\td{t}_{ji} \ln \ep \right ). \la{z1}
\ee
Note also that there are no finite terms in the boundary action when 
$g^0$ is flat.
The divergences can be cancelled by the covariant boundary action
\be
\CS_{\rm{ct}} = \frac{l}{ \k_5^2} \int d^4x \sqrt{h} \left (3 
+ {\textstyle{1 \over 8}} 
(T_{ii} - 6) + {\textstyle{1 \over 8}} (T_{ii} - 6) \ln\ep \right ), \la{p2}
\ee
where $T_{ij}$ is the induced scalar matrix on the boundary. For the specific
Coulomb branch flow considered in \cite{bzs}, this reduces to the 
counterterm action given there. 

These counterterms can be used to show that any BPS domain wall solution 
which uplifts to a distribution of D3-branes with zero dipole moment, 
has zero action but to do so
we must address two subtleties. The counterterm action removes all IR
divergences for such solutions - but there could still be finite contributions
to the action from both the boundary at infinity and the interior, usually
singular, boundary which need to be cancelled by additional finite 
counterterms. 

However, we can show that there are
no finite terms on the boundary at infinity. We said above that there are no
finite terms on the boundary from the bulk action (\re{z1}). 
There are also no
finite terms induced by the counterterm action, when we evaluate it
with the fields (\ref{x1}). From the arguments of the previous section
any brane solution will have to be brought into this coordinate frame to
use the (uplifted) renormalised action. There will be no finite terms
from the IR boundary for any BPS brane solution since it can be expressed 
exactly in the form (\ref{x1}).
In what follows we will show - 
from the ten-dimensional perspective - that there are no finite
contributions from the interior. So (\re{p2}) will ensure
that the action for all such BPS brane distributions is zero, as required by
supersymmetry. 

Given the asymptotic expansions of the five-dimensional fields we can use
the Kaluza-Klein ansatz to uplift to ten dimensions. This results 
in the following expansion for the metric 
\bea
ds_{10}^2 &=& \left (1 + \half \r^2 G + \half \r^2 \ln \r \td{G} + .. \right ) 
\left [ \frac{d\r^2}{\r^2} + (\frac{1}{\r^2} + .. ) dx^{\a}dx_{\a} \right] \nn \\
&& + \left (1 - \half \r^2 G - \half \r^2 \ln \r \td{G} \right ) 
\left (g^0_{ab} + \r^2 g^2_{ab} + \r^2 \ln \r \td{g}^2_{ab} \right ) 
dx^a dx^b; \la{g1} \\
g^2_{ab} &=& - \frac{1}{2} G_{;ab} - G g^{0}_{ab}; \hsp 
\td{g}^2_{ab} = - \frac{1}{2} \td{G}_{;ab} - \td{G} g^{0}_{ab}, \nn
\eea
where both $G$ and $\td{G}$ are $l=2$ harmonics on the five-sphere and are
related to the five-dimensional scalar matrices as 
\be
G = t_{ij} \m^{i} \m^{j}; \hsp \td{G} = \td{t}_{ij} \m^i \m^j.
\ee
By uplifting the fields we could formally write the counterterm action
in ten dimensions, to be used with the caveat that the ten-dimensional
fields must first be brought into the same coordinate frame as
in (\re{g1}). We would also have to fix a gauge for the $4$-form and bring
any solution into that same gauge and use the radial cutoff $\r = \ep$ to
regulate the boundary. 

\bigskip

Now we would like to use the five-dimensional counterterm action to 
renormalise the action for a particular distribution of D3-branes, namely
two separated stacks of branes. This will provide an explicit example
of how the covariant five-dimensional counterterm action can be used to
renormalise the ten-dimensional action. Such a distribution is described by a
harmonic function of the form
\be
D \propto \left ( 
\frac{p}{\left | {\bf{r}} - {\bf{a}} \right |^4} + \frac{q}{\left | {\bf{r}} 
- {\bf{b}} \right |^4} \right ),
\ee
corresponding to $p$ D3-branes placed at ${\bf{r} = \bf{a}}$ and $q$ D3-branes
placed at ${\bf{r} = \bf{b}}$. 
By shifting the coordinate system we can eliminate the dipole moment of
the charge distribution and bring the harmonic function into the form
\bea
D &=& \frac{1}{2l^4} \left ( \frac{1}{\left | {\bf{r}} - {\bf{a}} \right |^4} 
+ \frac{1}{\left | {\bf{r}} + {\bf{a}} \right |^4} \right ); \nn \\
&=& \frac{1}{l^4} \left ( \frac{1}{r^4} + \frac{a^2}{r^6} 
(\cos^2 \td{\theta} - {\textstyle{1 \over 6}}) + ... \right), \la{d2}
\eea
where for simplicity we have also assumed that the two stacks are equal 
in number and in the second line we have used explicit coordinates on $R^6$. 
Let us now evaluate the action for such a distribution. There are
three contributions to (\re{k1}) and (\re{k2}), from boundaries at the brane
locations and from the boundary at infinity. 

We deal with the latter first. The metric defined by (\re{d1}) and 
(\re{d2}) can be brought into the form (\re{g1}) by coordinate transformations
of the form 
\bea
r &=& \r^{-1}  + \frac{a^2}{6}(\cos^2 \theta - {\textstyle{1 \over 6}}) \r 
+ .. \nn \\
\td{\theta} &=& \theta + \frac{a^2}{3} \cos\theta \sin\theta \r^2 + ..., 
\eea
where 
\be
G = \frac{a^2}{3} (\cos^2 \theta - {\textstyle{1 \over 6}}),
\ee
which is manifestly an $l=2$ harmonic on the five-sphere. This function
corresponds to a five-dimensional scalar matrix of the form
\be
t_{11} = {\textstyle{5 \over 18}} a^2, 
\hsp t_{ij} = - {\textstyle{1 \over 18}} a^2 \d_{ij} \hsp \forall i \neq 1,
\ee
with an appropriate choice of $\m^i$,
which is manifestly traceless as required. Having matched the ten-dimensional
fields to the asymptotic expansion of the five-dimensional fields, we can
determine the IR boundary contribution to the action using the 
five-dimensional action and counterterms already determined. This vanishes as
we previously claimed; there is no finite contribution. 

All that remains is to show that there is no contribution to the action
from the interior (generically singular) boundaries. (A generic discrete 
distribution of branes will be singular at (all but one of) the brane 
locations \ci{kw}.) At each brane location we need to evaluate the term
\be
\d\CS_{10} = \frac{1}{8 \cdot 5! \k_{10}^2} \int d\Sigma^{\s} A^{\m\n\r\t} 
F_{\s\m\n\r\t} \propto \int (n \cdot \del (\ln D)) d\Sigma, \la{s1}
\ee
where $n$ is the normal to the boundary. Since there are by definition 
source terms for the harmonic function at the boundary 
the integrand will always be singular here. However 
the boundary must always have zero nine-volume and so the integral vanishes. 

Consider a general (continuous) distribution of D3-branes in $R^6$. Then
(\re{s1}) is defined by choosing a 5-surface which completely encloses
the distribution. The integral must be independent of deformations of
the 5-surface under which the distribution is still enclosed. Therefore 
we can always choose the 5-surface to be the outer boundary of the 
distribution itself. For example, suppose we have a ball distribution of
branes in $R^2$ contained within $x^2 + y^2 \le a^2$: then we can choose
the 5-surface to be $x^2 + y^2 = a^2$ times the origin in the remaining 
$R^4$. Now the 5-volume of this surface manifestly vanishes and so thus
will the integral $(\re{s1})$ even though $D$ is singular here. 

The only way we could get a finite volume $5$-surface would be to arrange the 
branes within a finite sized 
$5$-ball in $R^6$. However, this geometry is unphysical 
since it involves ``branes'' of negative charge and tension which are neither
D3-branes nor anti-D3-branes \ci{fgpw}. So we can exclude this 
geometry from our
analysis. For all other continuous or discrete D3-brane distributions there
can be no contribution from inner boundary terms. Of course similar
arguments are used to show that there are no contributions from Gibbons-Hawking
boundary terms either. This completes the
proof that the renormalised action is zero for all such
BPS D3-brane distributions.

\bigskip

The procedure for renormalising the ten-dimensional action
outlined above could be extended systematically 
to the most general asymptotically $AdS_5 \times S^5$ solutions. In principle
one could derive all the counterterms needed for $SO(6)$ gauged supergravity
in five dimensions. One could then use the Kaluza-Klein ansatz to uplift
the solution. By bringing a given ten-dimensional solution into this same
form by coordinate transformations, order 
by order in the expansion near infinity, we could identify the 
effective five-dimensional field expansions corresponding to the 
ten-dimensional solution and then use the {\it covariant} five-dimensional 
counterterm action to remove the IR divergences. 

\section*{Acknowledgements}

This work has been funded in part by St John's College, Cambridge and by the 
PPARC SPG programme ``String theory and realistic field theory'', 
PPA/G/S/1998/00613. It was also funded by the EU-RTN programme ``The quantum
structure of spacetime and the geometric nature of fundamental interactions.''


\begin{thebibliography}{99}

\bm{m} J. Maldacena, Adv. Theor. Math. Phys. {\bf 2} (1998) 231, 
[hep-th/9711200].

\bm{w} E. Witten, Adv. Theor. Math. Phys. {\bf 2} (1998) 253, 
[hep-th/9802150].

\bm{gkp} S. Gubser, I. Klebanov and A. Polyakov, Phys. Lett. {\bf B428} (1998) 
105, [hep-th/9802109].

\bm{a} O. Aharony, S. Gubser, J. Maldacena, H. Ooguri and Y. Oz, 
Phys. Rep. {\bf 323} (2000) 183, [hep-th/9905111].

\bm{bk} V. Balasubramanian and P. Kraus, Commun. Math. Phys. {\bf
    208} (1999) 413, [hep-th/9902121].

\bm{kls} P. Kraus, F. Larsen and R. Siebelink, Nucl. Phys. {\bf
    B563} (1999) 259, [hep-th/9906127].

\bm{ejm} R. Emparan, C. Johnson and R. Myers, Phys. Rev. {\bf
    D60} (1999) 104001, [hep-th/9903238].

\bm{hs} M. Henningson and K. Skenderis, JHEP {\bf 9807} (1998) 023,
  [hep-th/9806087; hep-th/9812032].

\bm{hss} S. de Haro, K. Skenderis and S. Solodukhin, Commun. Math. Phys. 
{\bf 217} (2001) 595, [hep-th/0002230].

\bm{s} K. Skenderis, Int. J. Mod. Phys. A {\bf 16}, 740 (2001), 
[hep-th/0010138].

\bm{bzs} M. Bianchi, D. Freedman and K. Skenderis, JHEP {\bf 0108} (2001) 041,
[hep-th/0105276].

\bm{mu} W. Mueck, hep-th/0105270.

\bm{gppz} L. Girardello, M. Petrini, M. Porrati and A. Zaffaroni,
  Nucl. Phys. {\bf B 569} (2000) 451, [hep-th/9909047].

\bm{ps} J. Polchinski and M. Strassler, hep-th/0003136.

\bm{pw} K. Pilch and N. Warner, hep-th/0006066. 

\bm{mn} J. Maldacena and C. Nunez, Phys. Rev. Lett. {\bf 86} (2001) 588, 
[hep-th/0008001]. 

\bm{t} M. Taylor-Robinson, hep-th/0001177.

\bm{clp} M. Cveti\v{c}, H. L\"{u} and C. Pope, Phys. Rev. {\bf D62} (2000) 
064028, [hep-th/0003286].

\bm{gl} C. Graham and J. Lee, Adv. Math. {\bf 87} (1991) 186.

\bm{fg} C. Fefferman and C. Graham, ``Conformal Invariants'', in 
 {\it Elie Cartan et les Math\'{e}matiques d'aujourd'hui}
 (Ast\'{e}risque, 1985) 95. 

\bm{hr} S. Hawking and S. Ross, Phys. Rev. D {\bf 52} (1995) 5865, 
[hep-th/9504019].

\bm{cg1} M. Cveti\v{c} and S. Gubser, JHEP {\bf 9904} (1999) 024, 
[hep-th/9902195].

\bm{cg2} M. Cveti\v{c} and S. Gubser, JHEP {\bf 9907} (1999) 010,
[hep-th/9903132].

\bibitem{p} R. Penrose and W. Rindler, {\it Spinors and Spacetime},
  volume 2, chapter 9 (Cambridge University Press, Cambridge, 1986). 

\bibitem{ht} M. Henneaux and C. Teitelboim, Phys. Lett. {\bf B142}
  (1984) 355.

\bibitem{bh} J. Brown and M. Henneaux, Commun. Math. Phys. {\bf 104} (1986) 
207.

\bibitem{cl} M. Cveti\v{c} and F. Larsen, Phys. Rev. Lett. {\bf 82} (1999) 484,
[hep-th/9805097].

\bibitem{clpst} M. Cveti\v{c}, H. L\"{u}, C. Pope, A. Sadrzadeh and T. Tran, 
Nucl. Phys. {\bf B586} (2000) 275, [hep-th/0003103].

\bibitem{nvv} H. Nastase and D. Vaman, Nucl. Phys. {\bf 583} (2000) 211, 
[hep-th/0002028]. 

\bibitem{kw} I. Klebanov and E. Witten, Nucl. Phys. {\bf B 556} (1999)
89, [hep-th/9905104].

\bibitem{fgpw} D. Freedman, S. Gubser, K. Pilch and N. Warner, 
JHEP {\bf 0007} (2000) 038, [hep-th/9906194].


\end{thebibliography}
\end{document}